# Rotator and extender ferroelectrics: Importance of the shear coefficient to the piezoelectric properties of domain-engineered crystals and ceramics

Matthew Davis[*], Marko Budimir, Dragan Damjanovic, and Nava Setter

Laboratory of Ceramics, Institute of Materials, Swiss Federal Institute of Technology - EPFL, Lausanne, Switzerland

**ABSTRACT**

The importance of a high shear coefficient $d_{15}$ (or $d_{24}$) to the piezoelectric properties of domain-engineered and polycrystalline ferroelectrics is discussed. The extent of polarization rotation, as a mechanism of piezoelectric response, is directly correlated to the shear coefficient. The terms "*rotator*" and "*extender*" are introduced to distinguish the contrasting behaviors of crystals such as *4mm* BaTiO$_3$ and PbTiO$_3$. In "rotator" ferroelectrics, where $d_{15}$ is high relative to the longitudinal coefficient $d_{33}$, polarization rotation is the dominant mechanism of piezoelectric response; the maximum longitudinal piezoelectric response is found *away* from the polar axis. In "extender" ferroelectrics, $d_{15}$ is low and the collinear effect dominates; the maximum piezoelectric response is found *along* the polar axis. A variety of *3m*, *mm2* and *4mm* ferroelectrics, with various crystal structures based on oxygen octahedra, are classified in this way. It is shown that the largest piezoelectric anisotropies $d_{15}/d_{33}$ are always found in *3m* crystals; this is a result of the intrinsic electrostrictive anisotropy of the constituent oxygen octahedra. Finally, for a given symmetry, the piezoelectric anisotropy increases close to ferroelectric-ferroelectric phase transitions; this includes morphotropic phase boundaries and temperature induced polymorphic transitions.

---

[*] email: matthew.davis@epfl.ch

## I. INTRODUCTION

Much work has gone into investigating the origins of the "giant" piezoelectric properties of lead based relaxor-ferroelectric single crystals, $(1-x)Pb(Mg_{1/3}Nb_{2/3})O_3$-$xPbTiO_3$ [PMN-xPT] and $(1-x)Pb(Zn_{1/3}Nb_{2/3})O_3$-$xPbTiO_3$ [PZN-xPT], since their "rediscovery[1]" in 1997. Specifically, attention has been centered on the importance of *polarization rotation*[2] in the presence of the monoclinic phases[3,4] recently discovered at their morphotropic phase boundary (MPB) regions.

In the monoclinic phases of PMN-xPT and PZN-xPT, as in that of $(1-x)PbZrO_3$-$xPbTiO_3$ [PZT][5], the polarization is not fixed along a crystallographic axis but can rotate freely within a single mirror plane. Thus, the $M_A$ ($M_B$) and $M_C$ monoclinic states (see figure 1), with $\{101\}_C$ and $\{010\}_C$ mirror planes, respectively, form "structural bridges[3]" between the $<111>_C$, $<101>_C$ and $<001>_C$ polar directions of the *3m* rhombohedral (R), *mm2* orthorhombic (O) and *4mm* tetragonal (T) phases typically observed in cubic perovskite ferroelectrics; the subscript 'C' means the direction is defined with respect to the cubic axes of the high temperature, parent phase. It has been suggested that *polarization rotation*, in the presence of one or more of these monoclinic phases, is responsible for the giant piezoelectric response of PMN-xPT and PZN-xPT single crystals when an electric field is applied in a non-polar, often $<001>_C$, direction[3]. Importantly, such a piezoelectric response, via rotation of the polar vector, is distinct from the collinear piezoelectric effect whereby the polar vector elongates (*polarization extension*) in the presence of an electric field applied along the polar direction. *In situ* diffraction studies have shown that in PZT ceramics with MPB compositions, the large piezoelectric response is due predominately to polarization rotation, rather than the elongation of the polar vector[6].

However, polarization rotation does not actually require the presence of a zero-field or spontaneously-formed monoclinic phase[2,7]. For any ferroelectric, when an electric field ***E*** is applied along a non-polar direction, the polar vector ***P*** will generally rotate towards the direction of applied field due to a biasing energy term in the free energy expansion $\Delta G = -\mathbf{P}\cdot\mathbf{E} = -PE\cos\theta$[8]. Indeed, in truly *3m* rhombohedral, *mm2* orthorhombic or *4mm* tetragonal phases, it can be shown that application of an electric field along a non-polar $<001>_C$, $<101>_C$ or $<111>_C$ type direction will lead to rotation of the polar vector in a $\{101\}_C$ ($M_A$ or $M_B$) or $\{010\}_C$ ($M_C$) "monoclinic plane[9]". Importantly, when this occurs the zero-field *3m*, *mm2* or *4mm* symmetry will be broken and monoclinic symmetry will result[7]. It is debatable whether the resultant, non-zero field structure constitutes a monoclinic "phase"[10]; likewise, it is debatable whether some residual monoclinic distortion remains in the crystal or is lost after removal of the field. Whether the monoclinic "phases" observed correspond to truly zero-field monoclinic phases or locally field-distorted versions of their R, O or T parents has been discussed at length elsewhere[4,11,12]. However, although polarization rotation does not require a monoclinic phase, it may be enhanced by some degree of "monoclinicity[13]".

In Section II, we will discuss explicitly how in *3m*, *mm2* and *4mm* ferroelectrics the extent of polarization rotation, when an electric field is applied away from the polar direction, is related to the piezoelectric shear coefficients $d_{15}$ and $d_{24}$, and more fundamentally to the transverse dielectric susceptibilities $\chi_{11}$ and $\chi_{22}$. In contrast, polarization extension and the collinear piezoelectric effect, as quantified by the longitudinal piezoelectric coefficient $d_{33}$, are directly related to the longitudinal susceptibility $\chi_{33}$. Therefore, the relative effects of polarization rotation and polarization extension, when a field is applied in a general, non-polar direction, will be related to the anisotropy factors $d_{15}/d_{33}$ and $\chi_{11}/\chi_{22}$.

In Section III, by comparing these factors for various ferroelectrics, we will introduce the concepts of "rotator" and "extender" ferroelectrics to differentiate two distinct classes of crystal. We will then show how the proximity of phase transitions between ferroelectric phases affects the propensity for polarization rotation and polarization extension. Finally, in Section IV, we will discuss how rotator and extender ferroelectrics behave in domain-engineered configurations.

## II. ROTATION VS. EXTENSION OF THE POLAR VECTOR

For any ferroelectric crystal, the extent of polarization rotation under an applied field is related to the shear deformation of the crystal lattice, as determined by the piezoelectric shear coefficients, e.g. $d_{15}$ and $d_{24}$, and more fundamentally to the transverse dielectric susceptibilities $\chi_{11}$ and $\chi_{22}$. In contrast, the relative extension of the polar vector under an applied field is related to the collinear piezoelectric effect, as quantified by the longitudinal piezoelectric coefficient $d_{33}$, and more fundamentally to the longitudinal dielectric susceptibility $\chi_{33}$.

Consider a ferroelectric crystal with polarization $\boldsymbol{P} = (0,0,P_3)$ defined with respect to its standard crystallographic, or principal, axes $\{x_1, x_2, x_3\}$. For *4mm*, *mm2* and *3m* symmetries, the dielectric susceptibility $\chi_{ij}$ will have non-zero components $\chi_{11}$, $\chi_{22}$ and $\chi_{33}$[14]. Application of an electric field $\boldsymbol{E} = (0,0,E_3)$ along the polar direction will lead to a lengthening of the polar vector (polarization extension) by an amount $\Delta \boldsymbol{P} = (0,0,\Delta P_3)$ where $\Delta P_3 = \varepsilon_0 \chi_{33} E_3$. Because the dielectric susceptibility matrix is diagonal, there is no change in the perpendicular components of the polar vector (no polarization rotation).

For a field $\boldsymbol{E}$ applied in an arbitrary direction, the change in the polarization vector is $\Delta P_i = \varepsilon_0 \chi_{ij} E_j$. The spontaneous strain[15,16] $S_{ij} = Q_{ijkl} P_k P_l$ becomes $S_{ij} + \Delta S_{ij} = Q_{ijkl}(P_k + \Delta P_k)(P_l + \Delta P_l)$, where $Q_{ijkl}$ are the electrostrictive coefficients of the parent phase[16] defined, here, with respect to the principal axes of the ferroelectric phase. Note that, although the electrostrictive coefficients are usually given (for convenience) with respect to the principal axes of the cubic parent phase, it is equally valid to define the electrostrictive tensor with respect to the principal axes of the ferroelectric phase. In matrix notation, $4Q_{1313} = Q_{55}$, $Q_{1133} = Q_{13}$ and $Q_{3333} = Q_{33}$.

It follows that the resultant, piezoelectric lattice strain $\Delta S_{ij}$ can be expressed as:

$$\Delta S_{ij} = E_p d_{pij} \approx Q_{ijkl}(\Delta P_k P_l + \Delta P_l P_k) \qquad [1]$$

Therefore, for a field applied along the polar direction, the purely longitudinal lattice strain (due to the collinear piezoelectric effect) will be given by:

$$\Delta S_{33} = E_3 d_{333} = 2 Q_{3333} \Delta P_3 P_3 \qquad [2]$$

such that:

$$\Delta P_3 = \frac{E_3 d_{333}}{2 P_3 Q_{3333}} \qquad [3]$$

Thus, the relative extension of the polar vector is given by:

$$\frac{\Delta P_3}{P_3} = \frac{E_3 d_{333}}{2 P_3^2 Q_{3333}} \qquad [4]$$

That is, at least at low fields where linearity holds, the extension of the polar vector is proportional to the longitudinal piezoelectric coefficient.

In contrast, if an electric field is applied perpendicular to the polar vector, the polar vector will rotate away from its original position by some angle $\theta$. For the special case of an electric field applied along the principal $x_1$ axis, there will be a change in polarization given by $\Delta \boldsymbol{P} = (\Delta P_1, 0, 0)$ where $\Delta P_1 = \varepsilon_0 \chi_{11} E_1$; there is no polarization extension ($\Delta P_3 = 0$). The corresponding lattice strain, the piezoelectric shear strain, will be given by:

$$\Delta S_{13} = E_1 d_{113} = 2 Q_{1313} \Delta P_1 P_3 \qquad [5]$$

such that:

$$\Delta P_1 = \frac{E_1 d_{113}}{2 P_3 Q_{1313}} \qquad [6]$$

The angle of polarization rotation is thus given by:

$$\tan \theta = \frac{\Delta P_1}{P_3} = \frac{E_1 d_{113}}{2 P_3^2 Q_{1313}} \qquad [7]$$

That is, rotation of the polar vector is directly proportional to the piezoelectric shear coefficient $d_{113}$ or, in matrix notation, $d_{15}$.

More fundamentally, the above breaks down to permittivity. From equations [2] and [5] we have[15]:

$$d_{333} = 2\varepsilon_0 \chi_{33} Q_{3333} P_3 \qquad [8]$$

$$d_{113} = 2\varepsilon_0 \chi_{11} Q_{1313} P_3 \qquad [9]$$

which are familiar equations of thermodynamic phenomenology[15]. Therefore, equations [4] and [7] become in their most obvious and fundamental form:

$$\frac{\Delta P_3}{P_3} = \frac{E_{33}\varepsilon_0 \chi_{33}}{P_3} \qquad [10]$$

and

$$\tan\theta = \frac{\Delta P_1}{P_3} = \frac{E_1 \varepsilon_0 \chi_{11}}{P_3} \qquad [11]$$

which are, of course, direct results of the relation $\Delta P_i = \varepsilon_0 \chi_{ij} E_j$.

In summary, polarization extension, when an electric field (or some component of an electric field) is applied along the polar vector, is directly related to the collinear piezoelectric effect and the coefficient $d_{33}$ ($\equiv d_{333}$). On the other hand, when an electric field (or some component of an electric field) is applied perpendicular to the polar vector, the resultant polarization rotation is directly related to the shear coefficients $d_{15}$ ($\equiv 2d_{113}$) and $d_{24}$ ($\equiv 2d_{223}$). Polarization extension will be strongest in materials with a large longitudinal coefficient $d_{33}$ due to a large, relative $\chi_{33}$. Polarization rotation will be strongest in materials with a large shear coefficient $d_{15}$ (or $d_{24}$) due to a large transverse susceptibility $\chi_{11}$. Therefore, when an electric field is applied in some arbitrary direction, the relative contributions to the piezoelectric response from polarization extension (due to any component of field along the polar axis) and polarization rotation (due to any perpendicular component) will be described by the ratios $d_{15}/d_{33}$ and $\chi_{11}/\chi_{33}$, the *piezoelectric* and *dielectric anisotropy factors*, respectively. Again, both are intrinsically related via equations [8] and [9].

## III. ROTATOR AND EXTENDER FERROELECTRICS

The dielectric susceptibilities (at constant stress) of various *4mm*, *mm2* and *3m* ferroelectric crystals are given, with respect to their principal crystallographic axes, in table I. Unless otherwise stated, the values correspond to those at room temperature and are experimentally derived. Values marked with an asterisk were calculated within the framework of 6th order Landau-Ginzburg-Devonshire (LGD) theory[15,17]; the coefficients used were taken from the articles by Haun et al. (PZT)[18] and Bell (BaTiO$_3$)[8]. Note that the value of $Q_{44}$ quoted by Bell[8] for BaTiO$_3$ should be 0.059 m$^4$/C$^2$. The values for PbTiO$_3$, taken from another article by Haun and coworkers[17], are also based on a phenomenological fitting of experimental data from ceramic samples. The dielectric anisotropy factor, i.e. the ratio of transverse and longitudinal susceptibilities $\chi_{11}/\chi_{33}$, is also given for each material.

The corresponding piezoelectric coefficients for each material are listed in table II. Again, they are experimentally derived unless otherwise marked. The piezoelectric anisotropy factor, defined as the ratio $d_{15}/d_{33}$, is also given. Notably, a second piezoelectric anisotropy factor $d_{24}/d_{33}$ could also be defined for the orthorhombic crystals, for which $d_{15} \neq d_{24}$. For simplicity, we shall restrict our discussion here to $d_{15}/d_{33}$; the relative importance of both $d_{15}$ and $d_{24}$ and their evolution with temperature in orthorhombic BaTiO$_3$ has been discussed elsewhere[19].

Of the ferroelectric materials listed, BaTiO$_3$, PbTiO$_3$, (1-x)PbZrO$_3$-xPbTiO$_3$ (PZT), KNbO$_3$, PMN-xPT and PZN-xPT are all perovskites[20]. BaTiO$_3$ transforms from its cubic paraelectric phase (point group *m3m*) to a *4mm* phase at 120°C. Upon further cooling, it undergoes successive transformations to *mm2* orthorhombic and *3m* rhombohedral phases at 5°C and -90°C, respectively[20]. KNbO$_3$ shows the same sequence of transformations but is *mm2* orthorhombic at room temperature; it transforms from its tetragonal phase at around 203 °C and has a Curie point of around 418°C[20,21]. In contrast, PbTiO$_3$ remains *4mm* tetragonal down to 0 K; it's Curie point is 490°C[22]. The (1-x)PbZrO$_3$-xPbTiO$_3$ (PZT or PZ-xPT) solid solution exhibits a morphotropic phase boundary between *3m* and *4mm* ferroelectric phases at a lead titanate content of around x = 48 mol.%[20]. As noted in the Introduction, similar morphotropic phase boundaries are observed in PZN-xPT and PMN-xPT at around x = 9 and x = 35 mol.%, respectively; monoclinic phases have been reported at the MPBs of PZT, PMN-xPT and PZN-xPT[3].

In contrast, K$_{2.9}$Li$_{1.6}$Nb$_{5.1}$O$_{15}$ (KLN)[23] and members of the Pb$_{1-x}$Ba$_x$Nb$_2$O$_6$ (PBN)[15] solid solution assume the tungsten-bronze structure. A morphotropic phase boundary, similar to that of PZT, is found between low barium content *mm2* and high barium content *4mm* ferroelectric phases in Pb$_{1-x}$Ba$_x$Nb$_2$O$_6$ at around x = 40 mol.%[24]. K$_{2.9}$Li$_{1.6}$Nb$_{5.1}$O$_{15}$ is *4mm* tetragonal at room temperature and has a Curie temperature of 405 °C[23]. Finally, LiNbO$_3$ and LiTaO$_3$ are isostructural[25]. Both are rhombohedral *3m* (space group *R3c*) at room temperature. LiTaO$_3$ has point group $\bar{3}m$ in its non-polar phase above its Curie point 650 °C, while LiNbO$_3$ remains ferroelectric to temperatures above 1200 °C[26]. The structures of LiNbO$_3$ and LiTaO$_3$ are described in detail elsewhere[26].

Notably, however, the perovskite, tungsten-bronze and LiNbO$_3$-type structures are all based on BO$_6$ oxygen octahedra; all the crystals considered in this paper are "oxygen-octahedra ferroelectrics[27]". As noted by Yamada, the electromechanical properties of such crystals will be related to those of an idealized oxygen-octahedron and, hence, to those of each other[27]. For all oxygen-octahedra ferroelectrics, it can be assumed that the piezoelectric effect in the ferroelectric phase is due to electrostriction of the high temperature, paraelectric phase biased by spontaneous polarization. Moreover, the piezoelectric properties of tungsten-bronze and LiNbO$_3$-type ferroelectrics can be predicted based on electrostrictive coefficients which assume *m3m* cubic symmetry in their high temperature, parent phases, as is the case for perovskites[27]. In the following, cubic, paraelectric symmetry will be assumed for the electrostrictive coefficients of all the ferroelectric crystals considered.

Comparison of the room temperature piezoelectric coefficients *d$_{ij}$* of two "classical" perovskite ferroelectrics, BaTiO$_3$ and PbTiO$_3$, reveals important differences between the two. As noted above, both have identical crystal structures and are *4mm* tetragonal at room temperature. However, as shown in table II, although their longitudinal $d_{33}$ and transverse $d_{31}$ coefficients differ by less than 40%, the shear coefficient $d_{15}$ of BaTiO$_3$ is around 10 times higher than that of PbTiO$_3$. The piezoelectric anisotropy, defined as the ratio of shear and longitudinal coefficients $d_{15}/d_{33}$, is around 6 in BaTiO$_3$ but only 0.7 in PbTiO$_3$. As has also been pointed out elsewhere[19], and as discussed in Section II, this is related to very different dielectric anisotropies. In PbTiO$_3$, at room temperature, the ratio of longitudinal and transverse susceptibilities $\chi_{11}/\chi_{33}$ is around 2; in BaTiO$_3$ it is 34.

If all the piezoelectric properties ($d_{ij}$) of a monodomain single crystal are known, we can calculate its piezoelectric properties along a general non-polar direction ($d_{ij}*$) via simple coordinate transforms[28]. For example, calculation of the longitudinal piezoelectric coefficient $d_{33}*$ as a function of orientation reveals much about the piezoelectric anisotropy of a material.

$d_{33}*$ has been calculated as a function of orientation for both room temperature BaTiO$_3$ and PbTiO$_3$ using the method described elsewhere[28,29] and the coefficients in table II. It is shown in figures 2 and 3(b) for PbTiO$_3$ and BaTiO$_3$, respectively. Most notably, $d_{33}*$ is maximum along the $[001]_c$ polar direction in PbTiO$_3$ (see figure 2); in contrast, the maximum value of $d_{33}*$ lies at an angle away from the $[101]_c$ polar direction in BaTiO$_3$ [fig. 3(a)]. As pointed out by Damjanovic et al., the maximum value of $d_{33}*$ for *4mm* BaTiO$_3$ lies close to the $[111]_c$ direction at room temperature[30].

For any *4mm* crystal, the longitudinal piezoelectric coefficient as a function of angle $\theta$ away from the polar axis is given by[11]:

$$d_{33}* = (\cos\theta \sin^2\theta)d_{31} + (\cos\theta \sin^2\theta)d_{15} + \cos^3\theta d_{33} \qquad [12]$$

The contribution from the transverse piezoelectric coefficient is usually negative and generally small. However, at *all* angles away from the polar direction ($\theta \neq 0$), $d_{33}*$ picks up a contribution from the shear coefficient $d_{15}$ but an increasingly small contribution from the longitudinal coefficient $d_{33}$. Therefore, wherever $d_{15}$ is large compared to $d_{33}$, there will be a propensity for a larger piezoelectric response $d_{33}^*$ *away* from the polar axis. This is indeed the case in room temperature BaTiO$_3$, but not so in PbTiO$_3$.

From now on, we will refer to any ferroelectric, like PbTiO$_3$, which has a piezoelectric anisotropy $d_{15}/d_{33}$ below some critical value[30], as an "extender"; from the above, its piezoelectric response when an electric field is applied in an arbitrary direction will be dominated by the collinear piezoelectric effect due to extension of the polar vector. As a consequence, its longitudinal coefficient $d_{33}^*$ will be highest *along* the polar direction. In contrast, we will refer to any ferroelectric like BaTiO$_3$, which has a piezoelectric anisotropy $d_{15}/d_{33}$ greater than some critical value[30], as a "rotator". In such a material, polarization rotation, as related to the shear piezoelectric effect, is the dominant mechanism of piezoelectric response. As a result, $d_{33}^*$ is highest *away* from the polar direction.

The critical value for $d_{15}/d_{33}$ is generally between 1 and 2 and can be derived analytically, for a given material, if the electrostrictive coefficients are known[30]. For a *4mm* crystal, using the condition that $\partial d_{33}^*/\partial\theta = 0$ at the angle of maximum $d_{33}^*$ (i.e. $\theta_{max}$), it can be shown that the largest piezoelectric response will be found away from the polar direction ($\theta_{max} \neq 0°$) whenever:

$$\frac{d_{15}}{d_{33}} > \left(\frac{3}{2} - \frac{Q_{1133}}{Q_{3333}}\right) \qquad [13]$$

For BaTiO$_3$, the critical value is calculated[30] to be 1.9. From the electrostrictive coefficients given by Haun and coworkers[17], it is around 1.8 for PbTiO$_3$. It is also around 1.8 for both PZ-60PT and PZ-90PT[18]. Using the electrostrictive coefficients for K$_{2.9}$Li$_{1.6}$Nb$_{5.1}$O$_{15}$ (KLN)[23], the critical value is 1.7. Assuming the electrostrictive coefficients do not change with temperature, the critical value for a given material will be a constant[30].

Calculation of the critical value is more complicated for *mm2* and *3m* crystals. For *mm2* symmetry, the piezoelectric coefficient in an arbitrary, non-polar direction $d_{33}{}^*$ is a function of two Euler angles[11,28] $\theta$ and $\phi$. It is given by[11]:

$$d_{33}{}^* = \cos\theta \sin^2\theta \sin^2\phi (d_{15} + d_{31})$$
$$+ \cos\theta \sin^2\theta \cos^2\phi (d_{24} + d_{32}) \qquad [14]$$
$$+ \cos^3\theta\, d_{33}$$

Thus, to calculate the critical value we first need to solve the two equations given by $\partial d_{33}{}^*/\partial\theta = 0$ and $\partial d_{33}{}^*/\partial\phi = 0$. It turns out that the condition for a maximum in piezoelectric response being away from the polar axis (i.e. a rotator ferroelectric) is the same as that for *4mm* crystals: i.e. that given in equation [13]. Care is needed, however, when calculating the critical value since the electrostrictive coefficients are usually quoted with respect to the principal axes of the cubic, parent phase (the pseudocubic axes). Thus we need to transform the quoted values to the principal axes of the *mm2* ferroelectric phase. With respect to the principal axes of the orthorhombic phase, the electrostrictive coefficients $Q_{3333}$ and $Q_{1133}$ are given by:

$$Q_{3333} = \frac{1}{2}\left(Q_{11}^C + Q_{12}^C\right) + \frac{1}{4}Q_{44}^C \qquad [15a]$$

and

$$Q_{1133} = \frac{1}{2}\left(Q_{11}^C + Q_{12}^C\right) - \frac{1}{4}Q_{44}^C \qquad [15b]$$

where $Q_{11}^C$, $Q_{12}^C$ and $Q_{44}^C$ are the electrostrictive coefficients with respect to the principal axes of the cubic, paraelectric phase.

In this way, the critical value is calculated to be 1.1 for *mm2* BaTiO$_3$ (assuming that the electrostrictive coefficients are independent of temperature[15,30]). For orthorhombic KNbO$_3$, using the electrostrictive coefficients given by Günter[31], the critical value is around 1.0. Finally, the electrostrictive coefficients are not known for PZN-9PT but we can assume[32] that they lie between the limiting values for pure

Pb(Zn$_{1/3}$Nb$_{2/3}$)O$_3$ (PZN) and PbTiO$_3$ (PT); using the coefficients quoted in Abe et al. transformed to the orthorhombic axes[32], both limiting critical values are close to 1.2.

For *3m* symmetry, the piezoelectric coefficient in an arbitrary direction is given by[11]:

$$d_{33}^* = \cos\theta \sin^2\theta \sin^2\phi (d_{15} + d_{31})$$
$$+ \cos^3\theta d_{33} \quad [16]$$
$$+ \sin^3\theta \cos\phi (3\sin^2\phi - \cos^2\phi) d_{22}$$

Note that when $\phi = 90°$ there is no contribution from $d_{22}$ and the condition again reduces to that for *4mm* crystals (eq. [12]). Thus, the critical value given by equation [12] is an upper limit for the condition that the maximum piezoelectric response lies away from the polar axis. In general, the $d_{22}$ coefficient also adds a tendency for a maximum in response away from the polar axis; that is, it generates a larger piezoelectric anisotropy than that resulting from the shear piezoelectric effect alone, and a stronger "rotator" quality in general.

For *3m* crystals, the quoted values of electrostrictive coefficient need to be transformed to the principal axes of the rhombohedral phase, as follows:

$$Q_{3333} = \frac{1}{3}\left(Q_{11}^C + 2Q_{12}^C + Q_{44}^C\right) \quad [17a]$$

and

$$Q_{1133} = \frac{1}{6}\left(2Q_{11}^C + 4Q_{12}^C - Q_{44}^C\right) \quad [17b]$$

In this way, the over-estimated critical value (ignoring the contribution from $d_{22}$) is calculated to be 1.6 for *3m* BaTiO$_3$. For PMN-33PT, the critical value is expected to lie between the values calculated for pure PbTiO$_3$ (PT) and Pb(Mg$_{1/3}$Nb$_{2/3}$)O$_3$ (PMN), using the relevant coefficients transformed to the rhombohedral axes[16,33]: that is, 1.5 and 1.3, respectively. For PZN-7PT, the critical value is expected to lie between 1.6 and 1.5. Using the normalized electrostrictive coefficients given by Yamada, assuming cubic symmetry[27], the over-estimated critical values are 1.5 and 1.4 for LiTaO$_3$ and LiNbO$_3$, respectively. Finally, the critical values are 1.6 and 1.7 for PZ-10PT and PZ-40PT, respectively.

We can now start to classify the ferroelectrics in tables I and II as either rotators or extenders. Although the electrostrictive coefficients are not quoted by Shrout et al. for the Pb$_{1-x}$Ba$_x$Nb$_2$O$_6$ tungsten bronzes, we might assume it is close to the value for another tungsten bronze[27], *4mm* Ba$_2$NaNb$_5$O$_{15}$, which is 1.7. In any case, we can note that the critical value is dominated in all cases by the geometrical component of equation [12], that is, $3/2 = 1.5$; therefore, it can always be expected to lie close to 1.5. As will be discussed below, none of the *3m* rhombohedral or *mm2* orthorhombic materials in table II can be classified as extenders; they are all rotators. In contrast, of the *4mm* tetragonal crystals, there are only two room temperature rotators, BaTiO$_3$ ($d_{15}/d_{33} = 6.0$) and Pb$_{0.56}$Ba$_{0.44}$Nb$_{2.00}$O$_6$ (2.1). The rest, including PbTiO$_3$ ($d_{15}/d_{33} = 0.7$), Pb$_{0.33}$Ba$_{0.70}$Nb$_{1.98}$O$_6$ (0.9) and the tetragonal PZT compositions are all extenders.

According to equations [8] and [9], the piezoelectric anisotropy factor $d_{15}/d_{33}$ is directly related to the dielectric anisotropy $\chi_{11}/\chi_{33}$ via the electrostrictive coefficients. The ratio between $d_{15}/d_{33}$ and $\chi_{11}/\chi_{33}$ will depend on the ratio of two electrostrictive coefficients; from equations [8] and [9]:

$$\left(\frac{d_{15}}{d_{33}}\right) = 2\frac{Q_{1313}}{Q_{3333}}\left(\frac{\chi_{11}}{\chi_{33}}\right) \qquad [18]$$

Again, the electrostrictive coefficients in equation [18] are given with respect to the principal axes of the relevant ferroelectric phase. For *4mm* crystals, the principal axes coincide with those of the cubic phase such that $4Q_{1313} = Q_{44}^C$ and $Q_{3333} = Q_{33}^C$. For *3m* crystals, the electrostrictive coefficient $Q_{1313}$ in the principal axis system is related to those in the cubic system by:

$$Q_{1313} = \frac{1}{6}\left(2Q_{11}^C - 2Q_{12}^C + \frac{1}{2}Q_{44}^C\right) \qquad [19]$$

For *mm2* crystals, the relevant equation is simply:

$$Q_{1313} = \frac{1}{2}\left(Q_{11}^C - Q_{12}^C\right) \quad [20]$$

For *4mm* PZ-60PT, PZ-90PT, PbTiO$_3$ and BaTiO$_3$, the ratios of electrostrictive coefficients ($2Q_{1313}/Q_{3333}$) are 0.41, 0.39, 0.38 and 0.27, respectively. For *mm2* KNbO$_3$ and BaTiO$_3$ they are 3.2 and 3.3, respectively. Lastly, for *3m* PZ-40PT, LiNbO$_3$, LiTaO$_3$, PZN and BaTiO$_3$, they are 2.8, 6.7, 4.9, 7.9 and 4.3, respectively.

$d_{15}/d_{33}$ and $\chi_{11}/\chi_{33}$ are plotted together in figure 4 for the three symmetry classes of materials listed in tables I and II. Various points can be made. Firstly, there is strong, linear correlation between $d_{15}/d_{33}$ and $\chi_{11}/\chi_{33}$ for crystals of the *same* point group symmetry (*3m*, *mm2* or *4mm*); the correlation does not depend on crystal structure (i.e. perovskite, tungsten bronze and LiNbO$_3$-type). Furthermore, the tendency for a larger piezoelectric anisotropy, for the same dielectric anisotropy, is highest in rhombohedral crystals and lowest in tetragonal crystals. Lines of best fit (forced through the origin) yield gradients of 0.17, 1.2 and 4.0 for the *4mm*, *mm2* and *3m* crystals, respectively. The relative trends in these values fit well with those for the theoretical values calculated above.

Why the ratio between $d_{15}/d_{33}$ and $\chi_{11}/\chi_{33}$ depends primarily on point group symmetry, and not on the crystal structure of the various oxygen-octahedra ferroelectrics considered, is a consequence of geometry. From the work of Yamada[27], we can assume that the ratios between electrostrictive coefficients will vary little between crystals with oxygen-octahedra based structures. Therefore, the differing gradients for the *3m*, *mm2* and *4mm* crystals will result from the relevant coordinate transformation of the ratio $2Q_{1313}/Q_{3333}$ required in each case (equations [15a, 17a, 19 and 20]). That is, the different correlations between the dielectric and piezoelectric anisotropies, $\chi_{11}/\chi_{33}$ and $d_{15}/d_{33}$, for the three symmetries result from different orientations of the polar axis with respect to the component oxygen octahedra; the three distinct correlations observed in figure 4 are a consequence of the intrinsic *m3m* anisotropy of the electrostrictive coefficients of the constituent oxygen octahedron. We can note that, for an isotropic material, there are two instead of three independent electrostrictive coefficients, since $Q_{44} = 2(Q_{11} - Q_{12})$[16]. If this condition is substituted into equations [15a, 17a, 19 and 20], the ratio $2Q_{1313}/Q_{3333}$ will become identical for all three symmetries. Finally, as noted by Yamada, the electrostrictive coefficients of all oxygen-octahedra ferroelectrics can be reduced to those of an ideal oxygen octahedra[27]. For these values ($Q_{11}^C = 0.10 \, m^4/C^2$, $Q_{12}^C = -0.034 \, m^4/C^2$ and $Q_{44}^C = 0.029 \, m^4/C^2$), the predicted ratios are 0.15, 3.3 and 4.6, for *4mm*, *mm2* and *3m* crystals, respectively.

Budimir and coworkers have recently shown that a large dielectric anisotropy, and hence a large piezoelectric anisotropy, is related to a proximity to ferroelectric-ferroelectric phase transitions whether induced by changes in composition or temperature, or by application of an electric field or stress[19]. In more fundamental terms, this can be seen to be due to a flattening of the free energy function when two or more ferroelectric phases become nearly degenerate[34]. Such free energy flattening also occurs when a field or stress is applied close to the coercive limit[34].

Therefore, the weak piezoelectric anisotropy of PbTiO$_3$ (table II) is explained by an absence of ferroelectric-ferroelectric phase transitions. As noted above, in contrast to BaTiO$_3$, PbTiO$_3$ remains tetragonal down to very low temperature. On the other hand, BaTiO$_3$ undergoes a phase transition to an *mm2* orthorhombic phase at around 0°C. The effect of the phase transition can be clearly seen by plotting the piezoelectric coefficients of the *4mm* phases of BaTiO$_3$ and PbTiO$_3$, calculated within the framework of 6th order LGD theory, as a function of temperature. These coefficients, taken from a previous publication[35], are shown in figure 5.

The effect of the proximity of the phase transition to an orthorhombic phase is clearly evident in figure 5(b). From table II, the ratio of $d_{15}/d_{33}$ is 6.0 for BaTiO$_3$ at room temperature but only 0.5 for BaTiO$_3$ at 380 K, far away from the phase transition. Thus, although BaTiO$_3$ is a rotator ferroelectric at room temperature, close to the transition to an orthorhombic phase, it becomes an extender ferroelectric at high temperatures. As a result the maximum piezoelectric response is observed along the polar axis in high temperature BaTiO$_3$; this is shown in figure 3(b). In contrast, PbTiO$_3$ is an extender at all temperatures. However, it is interesting to note that when destabilized by external, compressive stress, PbTiO$_3$ becomes a rotator[34].

The phenomenologically derived piezoelectric coefficients for BaTiO$_3$ across the entire temperature range, for *4mm*, *mm2* and *3m* phases, can be found in a previous publication[19]. Importantly, both the orthorhombic and rhombohedral phases of BaTiO$_3$ are rotators at *all* temperatures. As shown in table II, the piezoelectric anisotropy $d_{15}/d_{33}$ in the *mm2* phase at 225 K is 4.4; in rhombohedral BaTiO$_3$ at 0 K, it is around 7. This may be related to the fact that none of the *mm2* and *3m* ferroelectric crystals listed are extenders. Orthorhombic KNbO$_3$, for example, has a $d_{15}/d_{33}$ ratio of 7.0 at room temperature and is a strong rotator. As discussed elsewhere[19], in *mm2* BaTiO$_3$, $d_{15}/d_{33}$ increases toward the orthorhombic-tetragonal phase transition whereas $d_{24}/d_{33}$ increases toward the orthorhombic-rhombohedral phase transition. However, both ratios are always much greater than the critical value of 1.

An increase in piezoelectric and dielectric anisotropy is also observed close to morphotropic phase boundaries (MPBs), which constitute chemically-induced phase transitions. As shown in table II, and elsewhere[36], there is a clear increase in piezoelectric anisotropy close to the morphotropic phase boundary in both rhombohedral and tetragonal phases of PZT. In rhombohedral PZT, the rotator character is strongest near the MPB and decreases in compositions away from the MPB[37]. It can also be seen by comparing the piezoelectric coefficients of *4mm* PMN-38PT and PMN-42PT (table II). The ratio of $d_{15}/d_{33}$ is 1.3 in PMN-38PT but only 0.5 in PMN-42PT, further in composition from the MPB. Importantly, extremely large shear coefficients < 6000 pm/V are measured experimentally close to the MPB in PMN-xPT and PZN-xPT; as shown in table II, the piezoelectric anisotropy of rotator PZN-7PT is 65. Moreover, the effect is not unique to perovskites and is also observed, for example, in the Pb$_{1-x}$Ba$_x$Nb$_2$O$_6$ tungsten bronze system[24,38] (see table II). There, the increase in piezoelectric anisotropy close to the MPB occurs markedly in both orthorhombic and tetragonal phases[38]. Close to the MPB, *4mm* Pb$_{0.56}$Ba$_{0.44}$Nb$_{2.00}$O$_6$ is a rotator ($d_{15}/d_{33}$ = 2.1); in contrast *4mm* Pb$_{0.33}$Ba$_{0.70}$Nb$_{1.98}$O$_6$, away from the MPB, is an extender ($d_{15}/d_{33}$ = 0.9).

Finally, we can note from figure 5 that *all* piezoelectric coefficients increase near to the Curie point; such an increase corresponds to a dielectric softening of the material in all directions. This may have relevance to the abnormally large dielectric, piezoelectric and compliance coefficients observed in relaxor-ferroelectrics compared to other perovskite ferroelectrics[7,11]. The phase transition from the high temperature cubic, paraelectric phase is characteristically diffuse[39], or martensite-like, in relaxor-ferroelectrics like PMN-xPT and PZN-xPT with low PT contents. Therefore, the low temperature ferroelectric phase and high temperature paraelectric phase will remain close in energy over a large temperature range. Such dielectric softening may therefore be felt even at temperatures much lower than the temperature of peak permittivity.

**IV. DOMAIN ENGINEERED CRYSTALS AND CERAMICS**

Lastly, we can use the concepts of rotators and extenders to explain the piezoelectric properties of various domain-engineered and ceramic ferroelectrics. Assuming no extrinsic contribution to the response, the piezoelectric properties $d_{ij}^*$ of domain engineered crystals can be calculated if monodomain data measured along the polar direction (like that in table II) is available. Such calculations have recently been made by a variety of authors; details of the calculations can be found elsewhere[11,28].

Firstly, it has been shown both experimentally and by calculation that one can domain-engineer large, positive values of $d_{33}^*$ in room temperature BaTiO$_3$, KNbO$_3$, PMN-xPT and PZN-xPT[28,29,40,41]. As noted in the Introduction, "giant" piezoelectric coefficients are observed in $[001]_c$-poled rhombohedral or orthorhombic PMN-xPT and PZN-xPT. For example, the $d_{33}^*$ coefficient of $[001]_c$-poled PMN-33PT is around 2800 pm/V[42]; in contrast, in $[111]_c$-poled, *3m* rhombohedral PMN-33PT the longitudinal piezoelectric coefficient $d_{33}$ is only 190 pm/V (table II)[43]. As has been pointed out, this is predominantly due to the large piezoelectric anisotropy of PMN-33PT ($d_{15}/d_{33} = 22$)[29]. Similarly, although the longitudinal piezoelectric coefficient $d_{33}$ is only 90 along the $[001]_c$ polar axis of *4mm* BaTiO$_3$ (table II), $d_{33}^*$ coefficients of around 200 pm/V are measured in $[111]_c$-poled, domain-engineered crystals[40]. Again, this is due to the large piezoelectric anisotropy of rotator BaTiO$_3$ ($d_{15}/d_{33} = 6.0$). Similar enhancements in $d_{33}^*$ are observed in *mm2* KNbO$_3$ and PZN-9PT when poled along the non-polar $[001]_c$ direction[41,44].

Importantly, PMN-33PT, BaTiO$_3$, KNbO$_3$ and PZN-9PT are all rotators. In contrast, extender ferroelectrics like PbTiO$_3$, where the piezoelectric coefficient is highest along the polar axis (see figure 2), will not profit from domain engineering. $d_{33}$* can only be enhanced by domain-engineering in rotator ferroelectrics.

The transverse piezoelectric coefficient $d_{31}$* is also important to certain applications, such as piezoelectric benders, and should be minimized to reduced cross-talk in ultrasonic transducer arrays[28]. Thus, it is useful to know how the transverse piezoelectric coefficient might be tailored, for example, by domain engineering. In a previous paper we showed, by calculation, that large negative values of $d_{31}$* (> 1000 pm/V) can be predicted in $[001]_c$- and $[101]_c$-poled, domain-engineered PMN-33PT[28]. As we discuss, this is also a consequence of a strong piezoelectric anisotropy and a large shear coefficient. Similarly, enhanced transverse coefficients can be domain-engineered in BaTiO$_3$ and KNbO$_3$, as well[28]. Moreover, large transverse coefficients have also been measured experimentally in $[101]_c$-poled PMN-30PT[45]. The general result is, therefore, that large, negative values of $d_{31}$* can be domain-engineered in rotator ferroelectrics.

In contrast, however, there is a propensity for small and even positive values of $d_{31}$* in domain-engineered extender ferroelectrics like PbTiO$_3$ and BaTiO$_3$ at high temperatures. Whereas a $d_{31}$* of -190 pm/V is predicted for $[111]_c$-poled BaTiO$_3$, the equivalent value for $[111]_c$-poled PbTiO$_3$ is +4 pm/V[28]. As we discuss, the tendency for positive or zero transverse piezoelectric coefficients is not unique to polycrystalline ceramics, nor lead titanate[28,35].

## V. DISCUSSION AND CONCLUSIONS

The basic paradigm for the origin of the giant piezoelectric response of relaxor-ferroelectric single crystals PMN-xPT and PZN-xPT is that it results from polarization rotation in the presence of one or more monoclinic planes[3]. However, as pointed out in Section II, the concept of polarization rotation follows simply from the transverse component of dielectric susceptibility; that is, it does not require a monoclinic phase. Monoclinic symmetry will necessarily result when an electric field is applied to a *4mm* tetragonal, *mm2* orthorhombic or *3m* rhombohedral ferroelectric in a non-polar <001>$_C$, <101>$_C$ or <001>$_C$ direction. As pointed out by Kisi et al., the resultant, field-induced structure might be regarded as a monoclinic distortion rather than a true monoclinic phase[7,10].

Of course, the "giant" piezoelectric properties of PMN-xPT and PZN-xPT close to the MPB may indeed be due to true, zero-field monoclinic phases, the presence of which has been reported by many authors[3]; polarization rotation might be "easier" in zero-field monoclinic phases. As shown by Bell, an enhancement in piezoelectric activity can be predicted phenomenologically in monoclinic PZT over that of the rhombohedral and tetragonal phases[13]. Importantly, however, polarization rotation and monoclinic symmetry can also be expected in *4mm*, *mm2* or *3m* crystals when an electric field is applied in a non-polar direction; this includes the classical ferroelectrics $BaTiO_3$, $PbTiO_3$ and $KNbO_3$. Since the piezoelectric shear coefficients of the simpler perovskites are around an order of magnitude smaller than those of PMN-xPT and PZN-xPT (see table II), the resultant monoclinic shear distortions when a field is applied in a non-polar direction will be around 10 times smaller; they will therefore be difficult to resolve in diffraction experiments.

This article can be summarized as follows. Firstly, we have pointed out how the piezoelectric shear effect, as quantified by the coefficients $d_{15}$ and $d_{24}$, is related by electrostriction to polarization rotation, itself quantified by the transverse susceptibilities $\chi_{11}$ and $\chi_{22}$. In contrast, the collinear piezoelectric effect ($d_{33}$) is related to extension of the polar vector and the longitudinal susceptibility $\chi_{33}$. The piezoelectric and dielectric anisotropies, $d_{15}/d_{33}$ and $\chi_{15}/\chi_{33}$, are therefore directly correlated by the ratio of two electrostrictive coefficients. This correlation has been verified using both experimental and phenomenologically-derived data for a range of oxygen-octahedra ferroelectric crystals. From the range of data investigated, it seems that the correlation depends on point group symmetry only. Moreover, since the relative electrostrictive coefficients of all perovskite, tungsten-bronze and $LiNbO_3$-type structures are closely correlated[27], this is a result of the intrinsic, *m3m*, electrostrictive anisotropy of the constituent oxygen-octahedra. As a result, for a given dielectric anisotropy, piezoelectric anisotropy will be largest in *3m* crystals and smallest in *4mm* crystals.

Secondly, in crystals with large piezoelectric anisotropies ($d_{15}/d_{33}$), the piezoelectric response when a field is applied in a non-polar direction will be dominated by polarization rotation rather than polarization extension. The concept of "rotator" and "extender" ferroelectrics has been introduced. Rotator ferroelectrics have $d_{15}/d_{33}$ greater than a critical value and show enhanced piezoelectric properties along non-polar directions; the critical value lies close to 1.5 and is determined primarily by crystal symmetry. As a result they benefit from domain-engineering; large, positive longitudinal piezoelectric coefficients $d_{33}^*$ and large, negative transverse coefficients $d_{31}^*$ values can be domain-engineered in rotators such as PMN-33PT, $KNbO_3$ and $BaTiO_3$. In contrast, extender ferroelectrics like $PbTiO_3$ have $d_{15}/d_{33}$ ratios below this critical value; they show a maximum value of $d_{33}^*$ along the polar direction and will not generally benefit from domain-engineering. Domain-engineered, extender ferroelectrics show a propensity for small and even positive values of $d_{31}^*$.

Finally, a large piezoelectric anisotropy can be expected close to phase transitions between two or more ferroelectric phases. This includes those induced by changes in electric field, stress, temperature and composition. For example, *4mm* $BaTiO_3$ is a rotator at room temperature close to the transition to an *mm2* phase; in contrast, it becomes an extender at high temperatures. Morphotropic phase boundaries constitute chemically-induced phase transitions between ferroelectric phases. As a result, large piezoelectric anisotropies are observed close to the MPBs in PZT, PMN-xPT, PZN-xPT and the $Pb_{1-x}Ba_xNb_2O_6$ solid solution; the effect is not unique to perovskites.

**ACKNOWLEDGEMENTS**

The authors acknowledge financial support from the Swiss National Science Foundation.

# REFERENCES

1. S.-E. E. Park and T. R. Shrout, J. Appl. Phys. **82,** 1804-1811 (1997).
2. H. Fu and R. E. Cohen, Nature **403,** 281-283 (2000).
3. B. Noheda, Current Opinion in Solid State and Materials Science **6,** 27-34 (2002).
4. B. Noheda and D. E. Cox, Phase Transitions **79,** 5-20 (2005).
5. B. Noheda, D. E. Cox, G. Shirane, J. A. Gonzalo, L. E. Cross, and S.-E. Park, Appl. Phys. Lett. **74,** 2059-2061 (1999).
6. R. Guo, L. E. Cross, S.-E. Park, B. Noheda, D. E. Cox, and G. Shirane, Phys. Rev. Lett. **84,** 5423-5426 (2000).
7. E. H. Kisi, R. O. Piltz, J. S. Forrester, and C. J. Howard, J. Phys.: Condens. Matter **15,** 3631-3640 (2003).
8. A. J. Bell, J. Appl. Phys. **89,** 3907-3914 (2001).
9. D. Vanderbilt and M. H. Cohen, Phys. Rev. B **63,** 094108 (2001).
10. M. Davis, D. Damjanovic, and N. Setter, Phys. Rev. B **73,** 014115 (2006).
11. M. Davis, Thesis, Ecole Polytechnique Fédérale de Lausanne (EPFL), 2006.
12. M. Davis, Submitted to Journal of Electroceramics (2006).
13. A. J. Bell, J. Mat. Sci. **41,** 13-25 (2006).
14. J. F. Nye, *Physical properties of crystals*, 2nd ed. (Clarendon Press, Oxford, 1985).
15. M. J. Haun, E. Furman, S. J. Jang, and L. E. Cross, Ferroelectrics **99,** 13-25 (1989).
16. V. Sundar and R. E. Newnham, Ferroelectrics **135,** 431-446 (1992).
17. M. J. Haun, E. Furman, S. J. Jang, H. A. McKinstry, and L. E. Cross, J. Appl. Phys. **62,** 3331-3338 (1987).
18. M. J. Haun, Z. Q. Zhuang, E. Furman, S. J. Jang, and L. E. Cross, Ferroelectrics **99,** 45-54 (1989).
19. M. Budimir, D. Damjanovic, and N. Setter, J. Appl. Phys. **94,** 6753-6761 (2003).
20. B. Jaffe, W. R. Cook, and H. Jaffe, *Piezoelectric Ceramics* (Academic Press, New York, 1971).
21. J. Hirohashi, K. Yamada, H. Kamio, M. Uchida, and S. Shichiyo, J. Appl. Phys. **98,** 034107 (2005).

22  Z. Li, M. Grimsditch, X. Xu, and S.-K. Chan, Ferroelectrics **141,** 313-325 (1993).

23  M. Adachi and A. Kawabata, Jpn. J. Appl. Phys. **17,** 1969-1973 (1978).

24  T. R. Shrout, H. Chen, and L. E. Cross, Ferroelectrics **74,** 317-324 (1987).

25  T. Yamada, H. Iwasaki, and N. Niizeki, Jpn. J. Appl. Phys. **8,** 1127-1132 (1969).

26  R. S. Weis and T. K. Gaylord, Appl. Phys. A **37,** 191-203 (1985).

27  T. Yamada, J. Appl. Phys. **43,** 328-338 (1972).

28  M. Davis, D. Damjanovic, D. Hayem, and N. Setter, J. Appl. Phys. **98,** 014102 (2005).

29  D. Damjanovic, M. Budimir, M. Davis, and N. Setter, Appl. Phys. Lett. **83,** 527-529 (2003).

30  D. Damjanovic, F. Brem, and N. Setter, Appl. Phys. Lett. **80,** 652-654 (2002).

31  P. Gunter, Jpn. J. Appl. Phys. **16,** 1727-1728 (1977).

32  K. Abe, O. Furukawa, and H. Imagawa, Ferroelectrics **87,** 55-64 (1988).

33  J. Zhao, A. E. Glazounov, and Q. M. Zhang, Appl. Phys. Lett. **72,** 1048-1050 (1998).

34  M. Budimir, D. Damjanovic, and N. Setter, Phys. Rev. B **73,** 174106 (2006).

35  M. Davis, D. Damjanovic, and N. Setter, Appl. Phys. Lett. **87,** 102904 (2005).

36  M. J. Haun, E. Furman, S. J. Jang, and L. E. Cross, Ferroelectrics **99,** 63-86 (1989).

37  D. Damjanovic, J. Am. Ceram. Soc **88,** 2663-2676 (2005).

38  R. R. Neurgaonkar and W. K. Cory, J. Opt. Am. B **3,** 274-282 (1986).

39  G. Schmidt, Ferroelectrics **78,** 199-206 (1988).

40  S. Wada, S. Suzuki, T. Noma, T. Suzuki, M. Osada, M. Kakihana, S.-E. Park, L. E. Cross, and T. R. Shrout, Jpn. J. Appl. Phys. Pt. 1 **38,** 5505-5511 (1999).

41  K. Nakamura, T. Tokiwa, and Y. Kawamura, J. Appl. Phys. **91,** 9272-9276 (2002).

42  R. Zhang, B. Jiang, and W. Cao, J. Appl. Phys. **90,** 3471-3475 (2001).

43  R. Zhang, B. Jiang, and W. Cao, Appl. Phys. Lett. **82,** 787-789 (2003).

44  H. Dammak, A.-E. Renault, P. Gaucher, M. P. Thi, and G. Calvarin, Jpn. J. Appl. Phys. Pt. 1 **10,** 6477-6482 (2003).

45  J. Peng, H.-s. Luo, D. Lin, H.-q. Xu, T.-h. He, and W.-q. Jin, Appl. Phys. Lett. **85,** 6221-6223 (2004).

46  M. Zgonik, P. Bernasconi, M. Duelli, R. Schlesser, P. Günter, M. H. Garrett, D. Rytz, Y. Zhu, and X. Wu, Phys. Rev. B **50,** 5941-5949 (1994).

47  H. Cao and H. Luo, Ferroelectrics **274,** 309-315 (2002).

48  H. Cao, V. H. Schmidt, R. Zhang, W. Cao, and H. Luo, J. Appl. Phys. **96,** 549-554 (2004).

49  M. Zgonik, R. Schlesser, I. Biaggio, E. Voit, J. Tscherry, and P. Gunter, J. Appl. Phys. **74,** 1287-1297 (1993).

50  A. W. Warner, M. Onoe, and G. A. Coquin, J. Acoust. Soc. Am. **42,** 1223-1231 (1966).

51  R. Zhang and W. Cao, Appl. Phys. Lett. **85,** 6380-6382 (2004).

Table I

Dielectric susceptibilities for various oxygen-octahedra ferroelectrics with *3m*, *mm2* and *4mm* point group symmetries. Dependent quantities are shown in parentheses. Most data is experimentally derived and is taken from quoted values of unclamped relative permittivity ($\chi_{ij} = \varepsilon_{ij}^\sigma - 1$). The data marked with an asterisk is calculated according to 6th order Landau-Ginzburg-Devonshire (LGD) theory. All data corresponds to room temperature, unless noted otherwise.

| /[-] | $\chi_{11}$ | $\chi_{22}$ | $\chi_{33}$ | $\chi_{11}/\chi_{33}$ |
|---|---|---|---|---|
| Tetragonal *(4mm)* | | | | |
| BaTiO$_3$[46] | 4400 | (4400) | 130 | 34 |
| BaTiO$_3$ (380 K)* | 2030 | (2030) | 1030 | 2.0 |
| PbTiO$_3$[17]* | 124 | (124) | 67 | 1.9 |
| K$_{2.9}$Li$_{1.6}$Nb$_{5.1}$O$_{15}$ (KLN)[23] | 305 | (305) | 114 | 2.7 |
| PMN-38PT[47] | 4300 | (4300) | 733 | 5.9 |
| PMN-42PT[48] | 8630 | (8630) | 660 | 13 |
| Pb$_{0.33}$Ba$_{0.70}$Nb$_{1.98}$O$_6$[24] | 360 | (360) | 140 | 2.6 |
| Pb$_{0.56}$Ba$_{0.44}$Nb$_{2.00}$O$_6$[24] | 2500 | (350) | 350 | 7.1 |
| PZ-60PT* | 498 | (498) | 197 | 2.5 |
| PZ-90PT* | 120 | (120) | 72 | 1.7 |
| | | | | |
| Orthorhombic *(mm2)* | | | | |
| KNbO$_3$[49] | 149 | 984 | 43 | 3.5 |
| PZN-9PT[44] | 9000 | 21000 | 800 | 11 |
| Pb$_{0.86}$Ba$_{0.19}$Nb$_{1.98}$O$_6$[24] | 225 | 1200 | 1900 | 0.1 |
| BaTiO$_3$ (225 K)* | 498 | 2620 | 147 | 3.4 |
| | | | | |
| Rhombohedral *(3m)* | | | | |
| LiTaO$_3$[50] | 50 | (50) | 44 | 1.1 |
| LiNbO$_3$[50] | 83 | (83) | 29 | 2.9 |
| PMN-33PT[43,51] | 3950 | (3950) | 640 | 6.2 |
| PZN-7PT[(a)] | 11000 | (11000) | 700 | 16 |
| PZ-10PT* | 73 | (73) | 63 | 1.2 |
| PZ-40PT* | 120 | (120) | 98 | 1.2 |
| BaTiO$_3$ (0 K)* | 30 | (30) | 18 | 1.7 |
| | | | | |

(a) Presentation by K. Rajan et al. at the 2006 Navy Workshop on Acoustic Transduction Materials and Devices, State College, PA, USA

Table II

Piezoelectric coefficients for various oxygen-octahedra ferroelectrics with *3m*, *mm2* and *4mm* point group symmetries. Dependent quantities are shown in parentheses. Most data is experimentally derived. The data marked with an asterisk is calculated according to 6th order Landau-Ginzburg-Devonshire (LGD) theory. All data corresponds to room temperature, unless noted otherwise.

| /[pC/N] | $d_{31}$ | $d_{32}$ | $d_{33}$ | $d_{22}$ | $d_{24}$ | $d_{15}$ | $d_{15}/d_{33}$ |
|---|---|---|---|---|---|---|---|
| Tetragonal *(4mm)* | | | | | | | |
| $BaTiO_3$[46] | -33.4 | (-33.4) | 90 | 0 | (564) | 564 | 6.0 |
| $BaTiO_3$ (380 K)* | -185 | (-185) | 452 | 0 | (240) | 240 | 0.5 |
| $PbTiO_3$[17]* | -23.1 | (-23.1) | 79.1 | 0 | (56.1) | 56.1 | 0.7 |
| $K_{2.9}Li_{1.6}Nb_{5.1}O_{15}$ (KLN)[23] | -14 | (-14) | 57 | 0 | (68) | 68 | 1.2 |
| PMN-38PT[47] | -123 | (-123) | 300 | 0 | (380) | 380 | 1.3 |
| PMN-42PT[48] | -91 | (-91) | 260 | 0 | (131) | 131 | 0.5 |
| $Pb_{0.33}Ba_{0.70}Nb_{1.98}O_6$[24] | | | 60 | 0 | (52) | 52 | 0.9 |
| $Pb_{0.56}Ba_{0.44}Nb_{2.00}O_6$[24] | | | 185 | 0 | (380) | 380 | 2.1 |
| PZ-60PT* | -58.9 | (-58.9) | 162 | 0 | (169) | 169 | 1.0 |
| PZ-90PT* | -23.6 | (-23.6) | 80.0 | 0 | (51.8) | 51.8 | 0.6 |
| | | | | | | | |
| Orthorhombic *(mm2)* | | | | | | | |
| $KNbO_3$[49] | 9.8 | -19.5 | 29.3 | 0 | 156 | 206 | 7.0 |
| PZN-9PT[44] | 120 | -270 | 250 | 0 | 950 | 3200 | 13 |
| $Pb_{0.86}Ba_{0.19}Nb_{1.98}O_6$[24] | | | 70 | 0 | | <500 | <7.1 |
| $BaTiO_3$ (225 K)* | 19.8 | -50.1 | 52.6 | 0 | 447 | 233 | 4.4 |
| | | | | | | | |
| Rhombohedral *(3m)* | | | | | | | |
| $LiTaO_3$[50] | -2 | (-2) | 8 | 7 | (26) | 26 | 3.3 |
| $LiNbO_3$[50] | -1 | (-1) | 6 | 21 | (68) | 68 | 11 |
| PMN-33PT[43,51] | -90 | (-90) | 190 | 1340 | (4100) | 4100 | 22 |
| PZN-7PT[(a)] | -35 | (-35) | 93 | 1280 | (6000) | 6000 | 65 |
| PZ-10PT* | -1.1 | (-1.1) | 14.4 | 6.7 | (37.2) | 37.2 | 2.6 |
| PZ-40PT* | -5.2 | (-5.2) | 32.5 | 23.4 | (112) | 112 | 3.4 |
| $BaTiO_3$ (0 K)* | -0.5 | (-0.5) | 4.3 | 8.1 | (30.9) | 30.9 | 7.2 |

|  |  |  |  |  |  |  |  |
|--|--|--|--|--|--|--|--|

(a) Presentation by K. Rajan et al. at the 2006 Navy Workshop on Acoustic Transduction Materials and Devices, State College, PA, USA

**FIGURE CAPTIONS**

Fig. 1

Mirror planes of the monoclinic phases recently discovered at the morphotropic phase boundaries of PZT, PMN-xPT and PZN-xPT. The polar axes of the monoclinic phases are not fixed but can lie anywhere within their mirror plane between the limiting directions of the rhombohedral (R), orthorhombic (O) and tetragonal (T) phases. The notation is that after Vanderbilt and Cohen[9].

Fig. 2

$d_{33}$* [pm/V] as a function of orientation in (tetragonal) $PbTiO_3$ at room temperature. The highest value is found along the (vertical) polar direction; $PbTiO_3$ is an extender ferroelectric.

Fig. 3

$d_{33}$* [pm/V] as a function of orientation in (tetragonal) $BaTiO_3$ at (a) 298 K and (b) 380 K. At room temperature (a), the highest value is found *away from* the (vertical) polar direction: $BaTiO_3$ is a rotator ferroelectric. At 380 K (b), $BaTiO_3$ is an extender and the highest value of $d_{33}$* is found *along* the polar direction.

Fig. 4

Piezoelectric anisotropy versus dielectric anisotropy for the crystals listed in tables I and II, classified by their point group symmetry.

Fig. 5

(a) Piezoelectric coefficients of *4mm* $PbTiO_3$ as a function of temperature. $PbTiO_3$ is an "extender" at all temperatures. (b) Piezoelectric coefficients of *4mm* $BaTiO_3$ as a function of temperature. $BaTiO_3$ is a "rotator" at low temperatures close to the phase transition to an orthorhombic phase. However, it becomes an "extender" at high temperatures.

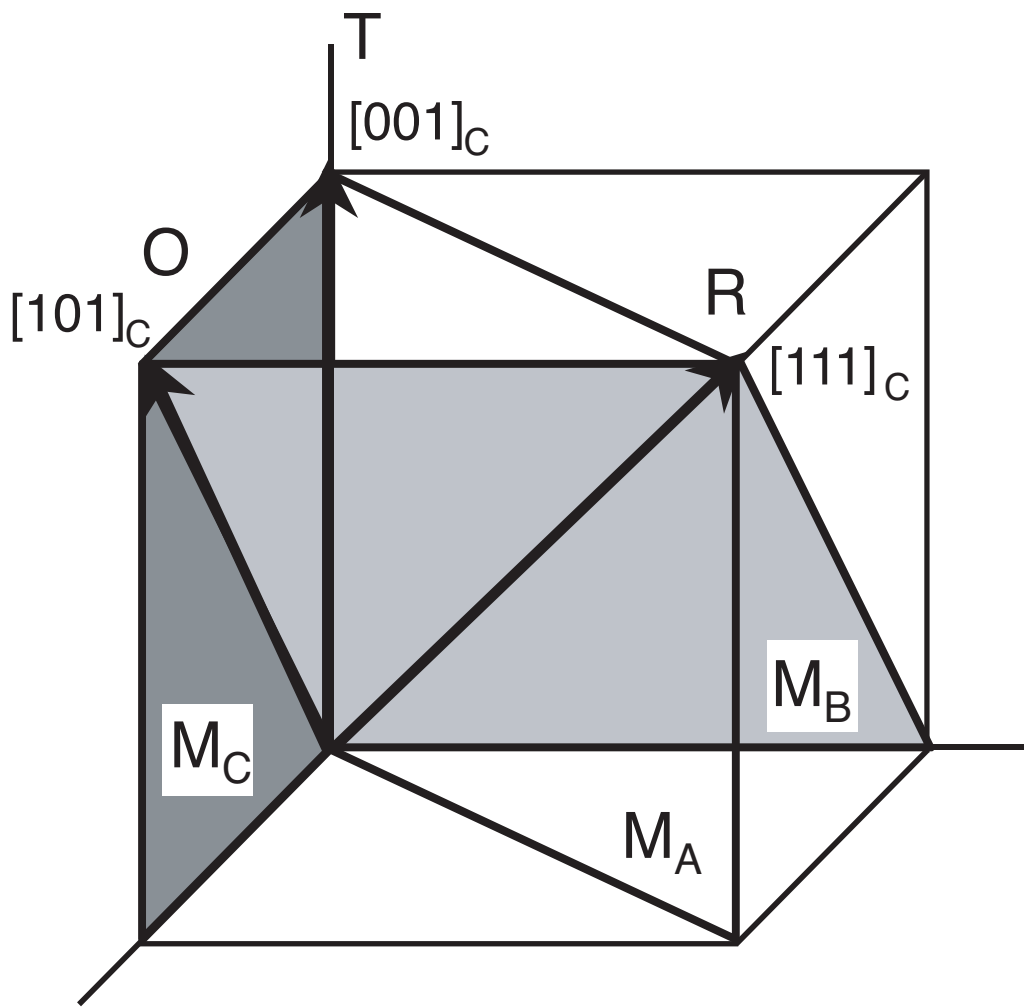

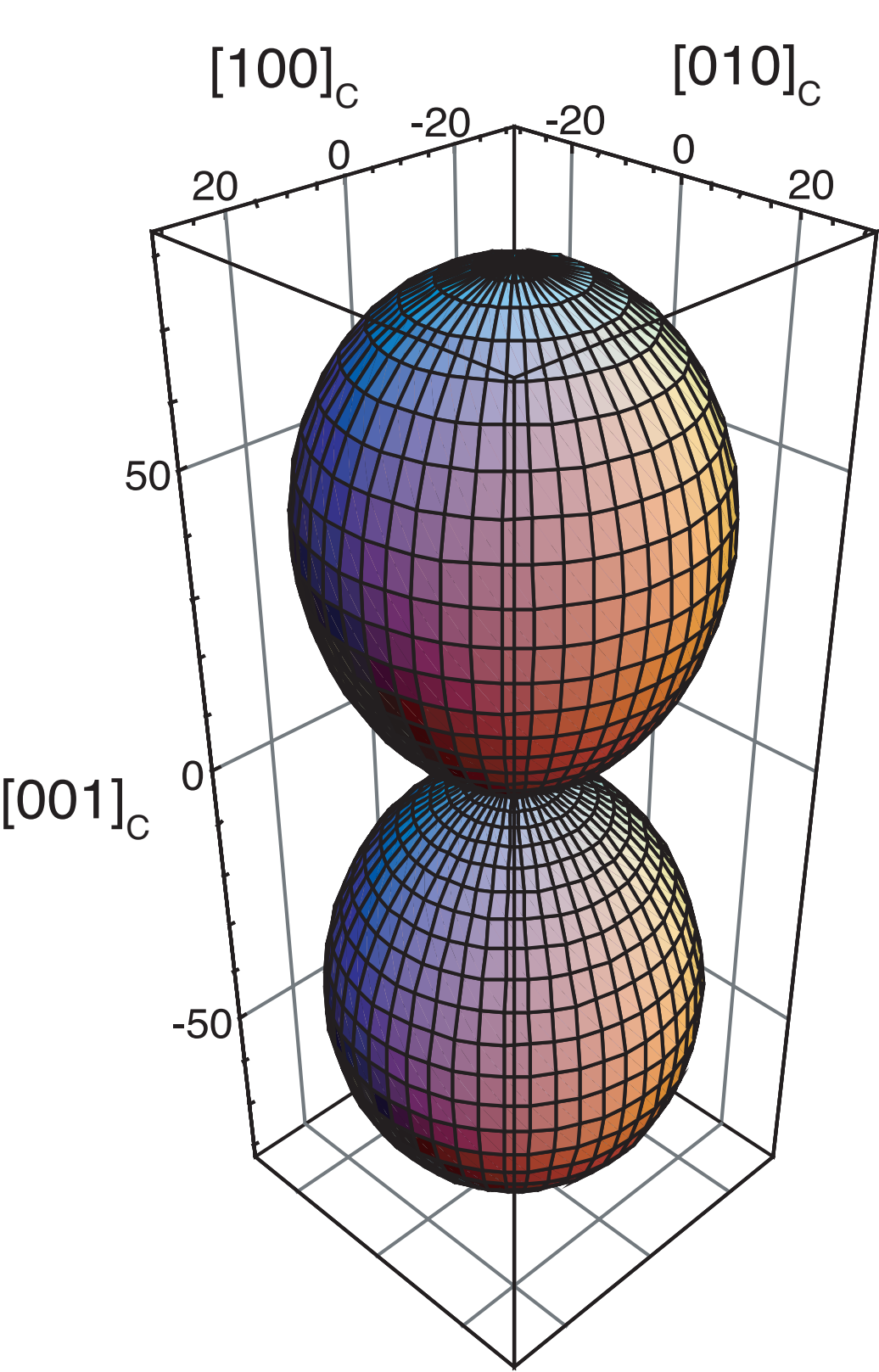

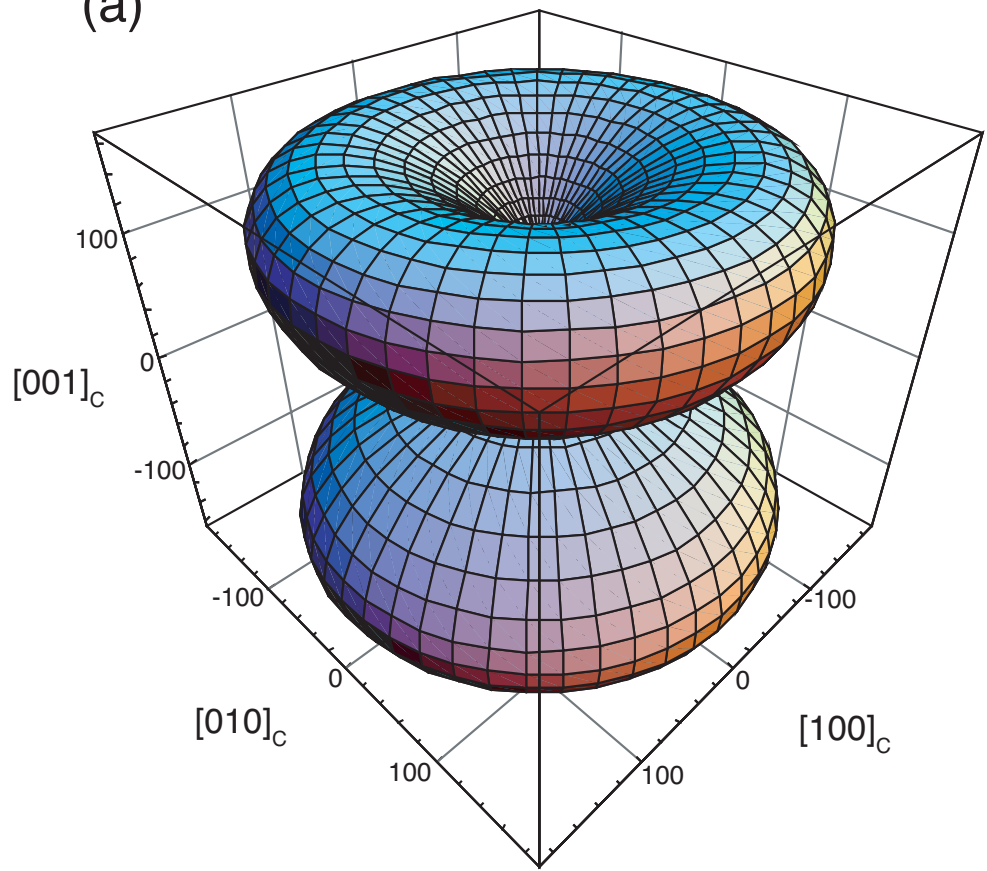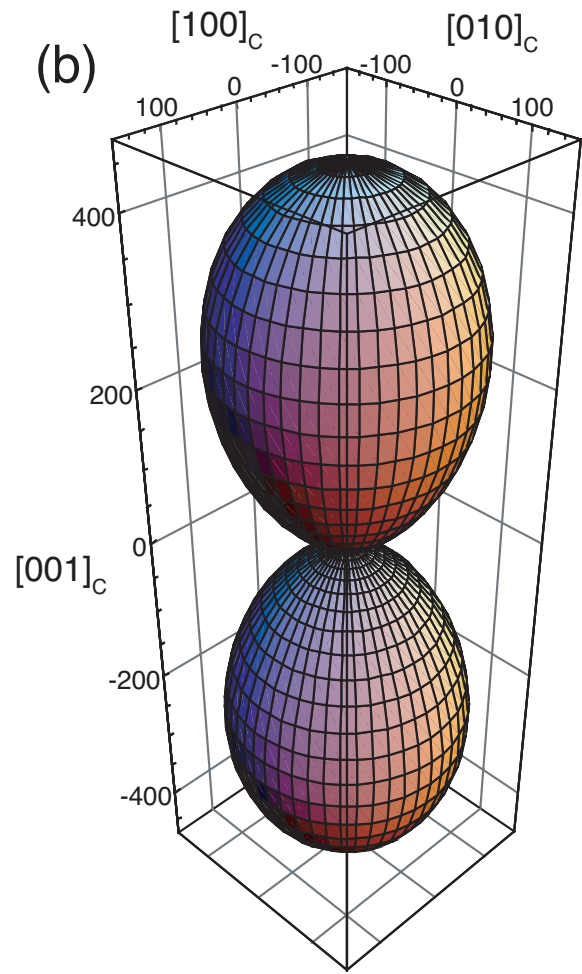

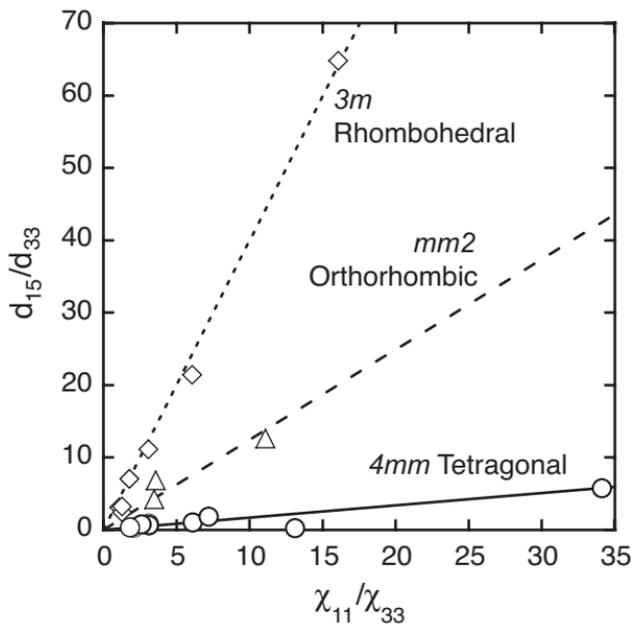

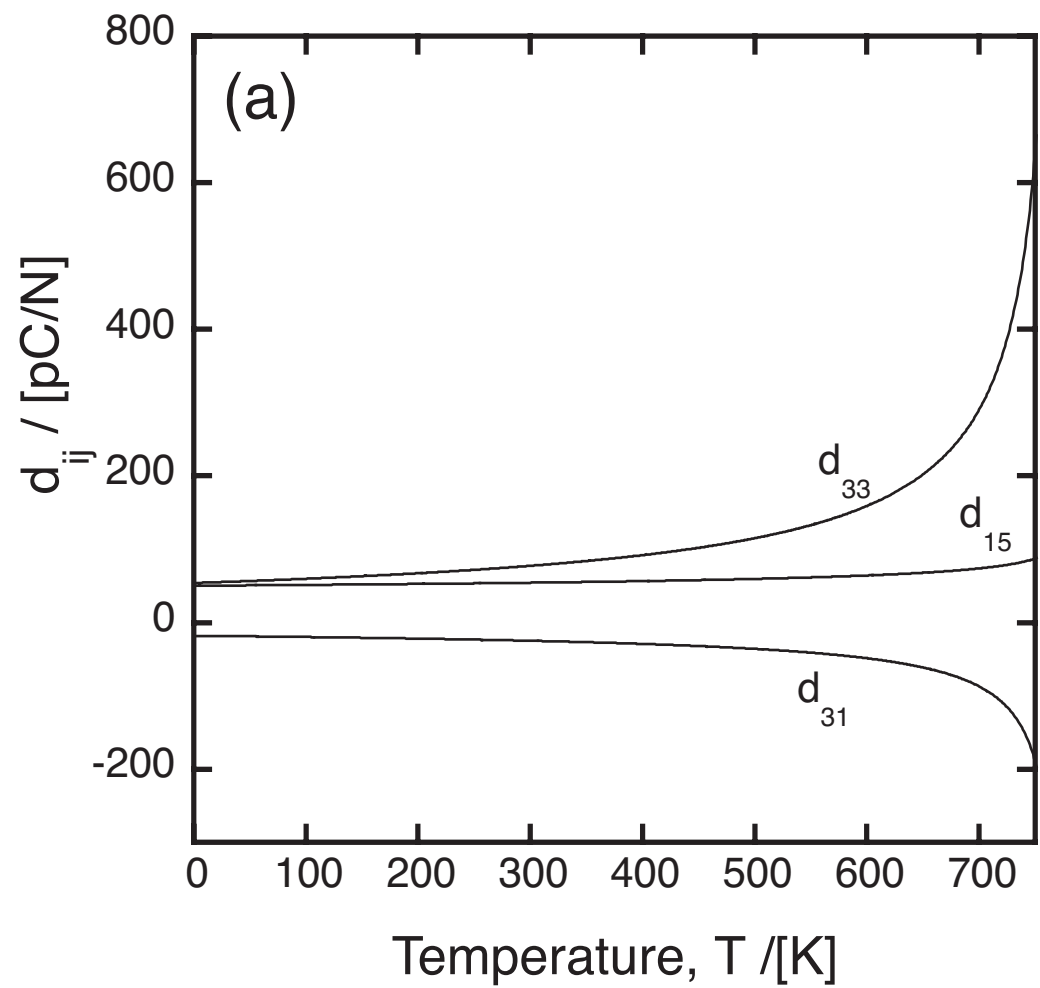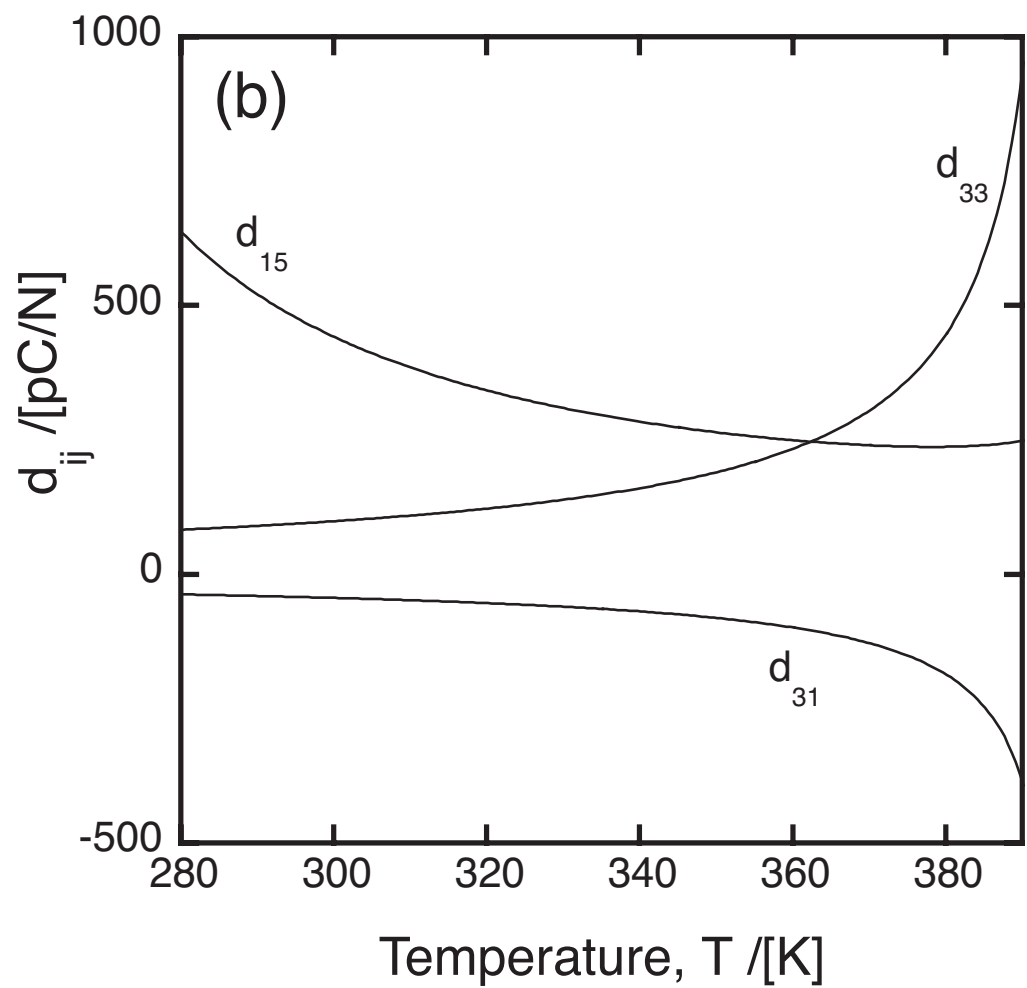